\documentclass{vow2008}

\usepackage{graphicx}

\title{GIPSY 3D: Analysis, visualization and VO-Tools}
\author{Ru\'iz, J. E.}
\author{Santander-Vela, J. D.}
\author{Espigares, V.}
\author{Verdes-Montenegro, L.}
\affil{Instituto de Astrof\'isica de Andaluc\'ia –- CSIC, Camino Bajo de Hu\'etor 50, 18008, Granada, Spain}
\author{van der Hulst, J. M.}
\affil{Kapteyn Astronomical Institute, Postbus 800, 9700 AV Groningen, The Netherlands}

\begin{document}

\maketitle


\begin{abstract}
The scientific goals of the AMIGA project are based on the analysis of a significant amount of spectroscopic 3D data. In order to perform this work we present an initiative to develop a new VO compliant package, including present core applications and tasks offered by the Groningen Image Processing System (GIPSY), and new ones based on \textit{use cases} elaborated in collaboration with advanced users. One of the main goals is to provide local interoperability between GIPSY (visualization and data analysis) and other VO software. The connectivity with the Virtual Observatory environment will provide general access to 3D data VO archives and services, maximizing the potential for scientific discovery.
\end{abstract}

\section{The AMIGA project}
The AMIGA (Analysis of the interstellar Medium of Isolated GAlaxies) project~\footnote{http://www.iaa.es/AMIGA.html} is an international scientific collaboration led from the Instituto de Astrof\'isica de Andaluc\'ia - CSIC. It focuses on a multiwavelenght analysis of a statistically significant sample of isolated galaxies, in order to provide a pattern of behaviour to the study of galaxies in denser environments. 

Since intensive analysis of 3D data is needed and given the experience acquired by the group in radio-VO developments~\citep{Ruiz:2009}, a collaboration has started with the Kapteyn Institute for upgrading the GIPSY software in order to produce a friendly VO-integrated package for high-level analysis of datacubes, with applicability to different multidimensional datasets and wavelengths. The final result will provide a more efficient way to treat data cubes, and will allow a way to get more and better science out of the data.

\section{GIPSY}
The Groningen Image Processing System~\citep[see][]{GIPSY:1992, GIPSY:2001}, developed at the Kapteyn Astronomical Institute, is one of the more mature and powerful systems available in order to analyze and visualize 3D data and to study the HI content in galaxies in particular.

3D datasets are the result of obtaining spectral information over a two-dimensional field of view. Present and future spectroscopic instrumentation, such as radio interferometers (including ALMA as well as other future radio facilities), Fabry Perot instruments and Integral Field Units in optical and NIR telescopes provides 3D information on gas and stars in galaxies. These complex data provide a wealth of information, but are at the same time much less exploited by the astronomical community due to their complexity.

A sophisticated image analysis system such as GIPSY is very well suited to deal with multi-dimensional datasets. Paradoxically the large number of parameters that can be fixed and/or determined is at the same time its power and its bane. Management and selection of such parameters is not sufficiently transparent and user-friendly that non-experts can fully benefit from the software and its full capability.

GIPSY needs the improvement of GUIs for user interaction with the data, detailed documentation and an easily installable and maintainable system. Preserving the core functionality (i.e. applications for modelling rotation curves, interactive inspection and characterisation of 3D data etc.) is of the utmost importance because this functionality is sparse and mostly absent in other image processing packages for astronomy.

The recently added Python bindings allow for the unique combination of existing high level core applications and state of the art new libraries which ensures an efficient path to improve graphical user interfaces, include VO functionality, and amend new data structures without the need to rewrite the core. 

\section{VO Tools}
In order to make GIPSY available to a larger scientific user base than the specialized radio astronomy community it is essential to make a proper connection to the VO environment. The VO is changing the paradigm for how astronomical data are exploited: not only are more data available in a uniform way, but also tools allowing exploitation of large volumes of data by astronomers from different communities are being delivered continually.

Interoperability is the \emph{Rosetta Stone} of the VO. It does not only allow concurrent access to distributed, heterogeneous data, but also enables VO software to communicate. The SAMP\footnote{Simple Access Message Protocol}~\citep{SAMP:2008} offers new born VO tools with all the functionalities coming from existing VO packages. Since they can communicate and share data sets, they form a huge VO meta-software in a continuously evolving ecosystem. Local interoperability with other VO software and access to VO archives will allow not only efficient multi-wavelength datasets comparisons but also the possibility to contribute to and benefit from this growing ecosystem of VO software, services and data.

\begin{figure}
\centering
\includegraphics[width=6cm]{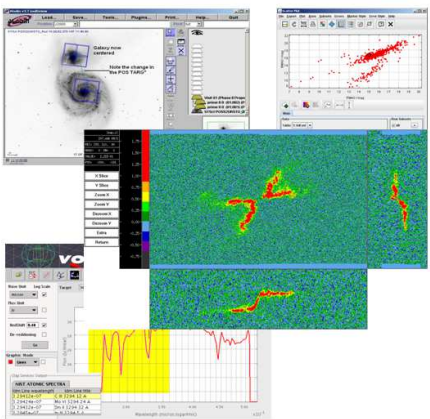}
\caption{GIPSY package will communicate with existing VO-tools thanks to local interoperability.
\label{fig:desktop}}
\end{figure}

\section{3D VO Archives and Services}

The increasing popularity of 3D data will result in an ever growing number of 3D VO archives. But because of the size and complexity of these datasets, data providers should supply on-line processing and analysis, and at the same time must have the capacity to store huge volumes of datacubes~\citep{Miller:2007}. Among the main issues and challenges that have to be addressed are: a data description model for datacubes in the VO, a storage format unification for 3D files and standard VO protocols for 3D discovery and data transmission. 

We will study the faisability of implementing GIPSY on the server side, providing VO services based on distributed computing analysis for a distributed storage 3D VO archive. The European OPTICON Network 3.6 and the NVO~\footnote{National Virtual Observatory} project in the US, with participation of other several groups, have provided a common desktop and server framework to bring data processing and analysis software into a common environment~\citep{Tody:2006}. As the GIPSY 3D project shares its philosophy, it must be fully compliant with this environment and must be taken into account from the very beginning.

\section*{Acknowledgments}

This work will be partially supported by the Spanish Ministry of Science and Innovation thanks to DGI Grant AYA2008-06181-C02-01.

\end{document}